\documentclass[prb,preprint,amsmath,amssymb,superscriptaddress]{revtex4-1}
\usepackage{graphicx}% Include figure files
\usepackage{dcolumn}% Align table columns on decimal point
\usepackage{hhline}
\usepackage{bm}% bold math

\usepackage{float}

\usepackage{amsmath}

\usepackage{epstopdf}

\begin{document}

%\wideabs{

\title{Evidence for Dimer Crystal Melting in the Frustrated Spin-Ladder  BiCu$_2$PO$_6$}

\author{K.-Y. Choi}
\email[]{kchoi@cau.ac.kr}
\affiliation{Department of Physics, Chung-Ang University, 221 Huksuk-Dong, Seoul 156-756, Republic of Korea}

\author{J.~W. Hwang}
\affiliation{Department of Physics, Chung-Ang University, 221 Huksuk-Dong, Seoul 156-756, Republic of Korea}

\author{P. Lemmens}
\affiliation{Institute for Condensed Matter Physics, TU Braunschweig, D-38106 Braunschweig, Germany}

\author{D. Wulferding}
\affiliation{Institute for Condensed Matter Physics, TU Braunschweig, D-38106 Braunschweig, Germany}

\author{G.~J. Shu}
\affiliation{Center for Condensed Matter
Sciences, National Taiwan University, Taipei 10617, Taiwan}

\author{F.~C. Chou}
\affiliation{Center for Condensed Matter
Sciences, National Taiwan University, Taipei 10617, Taiwan}

%\date{\today}

\begin{abstract}
In the spin ladder compound BiCu$_2$PO$_6$ there exists a decisive dynamics of spin excitations that we classify and characterize using inelastic light scattering. We observe low-energy singlets and a broad triplon continuum extending from 36~cm$^{-1}$ to 700~cm$^{-1}$ in ($aa$), ($bb$), and ($cc$) light scattering polarizations. Though isolated spin ladder physics can roughly account for the observed excitations at high energies, frustration and interladder interactions need to be considered to fully describe the spectral distribution and scattering selection rules at low and intermediate energies. More significantly, an interladder singlet bound mode at 24~cm$^{-1}$, lying below the continuum, shows its largest scattering intensity in interladder ($ab$) polarization. In contrast, two intraladder bound states  at 62~cm$^{-1}$ and  108~cm$^{-1}$  with energies comparable to the continuum are observed with light polarization along the leg ($bb$) and the rung ($cc$). We attribute the rich spectrum of singlet bound modes to a melting of a dimer crystal. Our study provides evidence for a Z$_2$ quantum phase transition from a dimer to a resonating valence bond state driven by singlet fluctuations.
\end{abstract}

\maketitle

Quantum spin ladders provide exceptional model systems, which enable us to explore rich one-dimensional physics, novel states of matter, and complex phase diagrams.~\cite{Dagotto,Dagotto96,Klanjsek}
The two-leg ladder is described by the leg ($J_\|$) and rung ($J_\perp$) couplings. For $r=J_\perp/J_\|\neq 0$, the ground state is given by a short-range resonating valence bond (RVB) state and elementary excitations are either dressed triplet excitations (triplons) or pairs of bound spinons with a finite gap.~\cite{White}

Spin ladders have shown a variety of fascinating phenomena and turned out to have a relevance to
diverse fields. In an applied field, they realize Tomonaga Luttinger liquids characterized by fractional $S=1/2$ spinon excitations.~\cite{Thielemann,Hong,Nafradi,Mukhopadhyay,Ninios,Schmidiger} Weakly-coupled spin ladders demonstrate confinements of spinon excitations, which are a condensed-matter analogue of the confinement of quarks in particle physics.~\cite{Lake}  The spin ladders show macroscopic quantum coherence of entangled spin pairs, holding a promise for quantum computation.~\cite{Li,Lorenzo} In hole-doped spin ladders, d-wave superconductivity has been reported and they provide a conceptual framework to understand unconventional superconductors.~\cite{Uhrig04}

Until now, isolated spin-ladder physics has been in the focus of research. To our best knowledge, the possibility of a quantum phase transition has not yet been experimentally addressed due to a lack of relevant materials. The spin ladder geometry may lead to two different routes to such transitions: Interladder couplings can induce a transition from a RVB state to a 2D N\'{e}el ordered state with classical spin-wave excitations,~\cite{Sachdev} while modifications of a leg to a frustrated  $J_1-J_2$ chain
can tune a ground state as well. When $J_\perp=0$ and $J_2\geq 0.24117 J_1$, two decoupled, dimerized chains are recovered. Upon turning on the rung coupling, RVB fluctuations are enhanced and eventually the dimer order melts into the RVB phase.~\cite{Lavarelo} BiCu$_2$PO$_6$ is an enticing material to address the crossover of a $J_1-J_2$ chain to a spin ladder since it realizes a two-leg ladder with frustrating couplings along the legs.~\cite{Tsirlin}

The spin topology of BiCu$_2$PO$_6$  is sketched in Fig.~1(a).
The crystal structure of BiCu$_2$PO$_6$ consists of complex ribbons with the basic unit being a folded pair of edge-sharing CuO$_4$ plaquettes. The neighboring plaquette pairs are connected by shared corners along the $b$ axis while the next-nearest-neighbor plaquettes are linked by PO$_4$ tetrahedra. The leg coupling  $J_\|=J_1$ appears between the corner-sharing plaquettes while the rung coupling of $J_{\bot}=J_4$ involves the plaquettes of neighboring ribbons. The frustrating next-nearest-neighbor couplings $J_2$ and $J'_2$ along the legs run between the copper plaquettes, joined by another plaquette via a PO$_4$ tetrahedron. Finally, the sizable interladder coupling $J_3$ is formed between the edge-sharing plaquettes. The static magnetic susceptibility, the magnetic specific heat, the Knight shift, and the magnetization evidence a non-magnetic singlet ground state with a spin gap of $\Delta=32~$K.~\cite{Tsirlin,Koteswararao,Koteswararao00,Bobroff,Alexander,Casola,Mentre,Kohama}
Geometrical frustration and anisotropic exchange and interladder interactions are invoked
to explain the field-induced transitions between different magnetic phases, the reduction of the spin gap, and the impurity-induced magnetic ordering.~\cite{Tsirlin,Bobroff,Kohama} In particular, the field-induced incommensurate
magnetic ordering is beyond simple spin ladder physics. However, there exists no experimental knowledge about the possibility of a Z$_2$ transition from a columnar dimer phase (with a translational symmetry) to a RVB one (without a translational symmetry).
A hallmark of such a transition is the presence of softening singlet modes.~\cite{Vekua} Raman spectroscopy is a unique experimental method which can probe magnetic excitations in the singlet sector, information which is essential to understand the role of low-energy singlet fluctuations in driving a transition between two different singlet states.~\cite{Lemmens}

In this paper we present magnetic Raman scattering of BiCu$_2$PO$_6$ as a function of temperature and polarization. The most distinct feature is the presence of singlet bound states of two triplons in both {\it interladder} and {\it intraladder} polarizations. The intraladder singlet bound modes are the anticipated melting of a dimer phase through singlet fluctuations
while the interladder singlet bound mode heralds the transition to a 3D magnetically ordered state. This demonstrates that
BiCu$_2$PO$_6$ is in the vicinity of two quantum critical points while being at the border line to dimer and 3D ordered phases.

Single crystals of BiCu$_2$PO$_6$ were grown by floating-zone method. The dimensions of the investigated single crystal is $4\times 2\times 1 ~\mbox{mm}^3$. The sample was kept in an evacuated optical closed-cycle cryostat with a temperature range from room temperature down to 3~K. Raman scattering measurements were carried out in quasi-backscattering geometry with the excitation line $\lambda =532$~nm of a Nd:yttrium aluminum garnet (YAG) solid-state laser. The incident power of 3~mW was focused onto the crystal surface with a spot diameter of about 100 $\mu$m to avoid heating effects. To fully suppress Rayleigh and stray light scattering, the low-energy cutoff of Raman spectra was set to $\omega=10~\mbox{cm}^{-1}$. Raman spectra were collected via a DILOR-XY triple spectrometer and a nitrogen-cooled charge-coupled device detector.

\begin{figure}[htp]
\centering
\includegraphics[width=10cm]{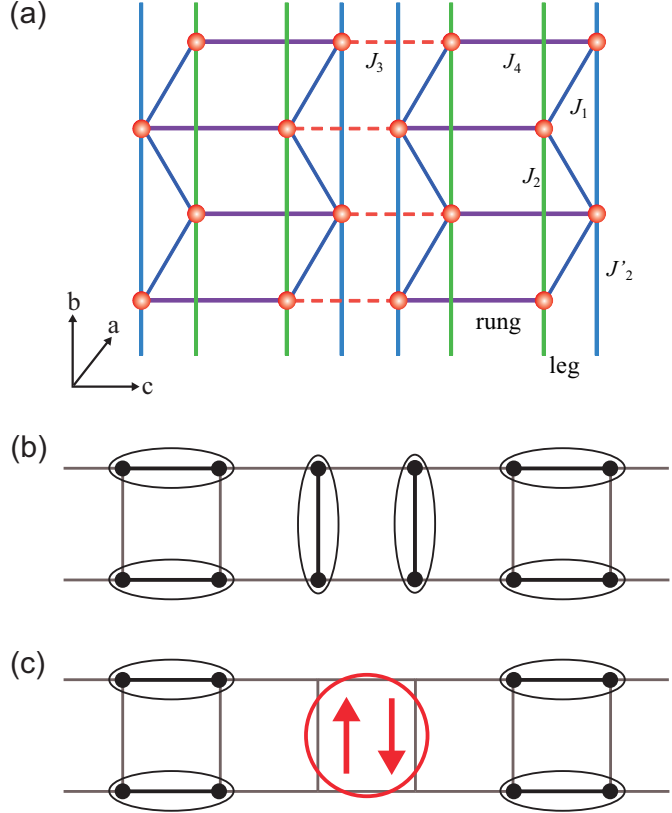}
\caption{(Color online)(a) Schematic of a frustrated spin ladder with
  the nearest neighbor leg coupling $J_1$, the rung coupling $J_4$, and the frustrating
  next-nearest-neighbor leg couplings $J_2$ and $J'_2$. $J_3$ denotes the interladder coupling.
  (b) and (c) Cartoons of two-triplets Raman scattering in a RVB picture. (b) a snapshot of
  the ground state; (c) locally excited singlet state of two neighboring triplet excitations depicted
  as a pair of the opposite arrows.}
\label{FIG1}
\end{figure}

Figure~2(a) shows Raman spectra at 3 and 280~K obtained in ($aa$), ($bb$), and ($cc$) polarization, respectively.
We identify 56 out of 60 Raman-active phonon modes as predicted from the factor group analysis ($Pnma$ space group);
$\Gamma_{Raman}= 18A_g(aa, bb, cc)+12B_{1g}(ab)+18B_{2g}(bc)+12B_{3g}(ca)$. Upon cooling from 280~K to 30 K the phonon modes show a moderate hardening due to lattice anharmonicity.
For temperatures below $\Delta=32$~K, a slight softening of the phonon frequency by 0.2-0.4 cm$^{-1}$ is observed (not shown here). Since such spin-phonon coupling is a general feature of low-dimensional antiferromagnets,~\cite{Choi03} we omit a further discussion of the phonon modes.

Hereafter, our analysis focuses on magnetic excitations. For spin ladders, inelastic scattering of singlets by light occurs as simultaneous excitations of two neighboring singlets into a higher singlet state of two bound triplets
since  a Raman scattering process conserves a total spin of $z$-component ($S_z^{tot}=0$). This process is sketched
in Figs.~1(b) and (c). Figure~1(b) illustrates a snapshot of a nonmagnetic spin singlet of two S=1/2 Cu$^{2+}$ spins. Low-energy  excitations are composite bosons of two-particle states, which
are a linear combination of triplets $\{|t^0\rangle, |t^1\rangle, |t^{-1}\rangle\}=\{|\uparrow\downarrow \rangle+|\downarrow\uparrow \rangle, |\uparrow\uparrow\rangle, |\downarrow\downarrow\rangle\}$.
The final state of Raman scattering thus corresponds to $|t_i^0\rangle |t_{i+1}^0\rangle+ |t_i^1\rangle |t_{i+1}^{-1}\rangle+|t_{i}^{-1}\rangle|t_{i+1}^1\rangle$. This two-triplet excitation is drawn
as a pair of opposite arrows in Fig.~1(c). Here the arrows represent the excited triplets. If there
is an attractive interaction between the excited triplets, a singlet bound state ($t^{1}t^{-1}$ or $t^{0}t^{0}$) is formed in the similar way
as mesons ($q\bar{q}$) are created by the confinement of quark and antiquark.

In the following, we introduce a polarization notation specific to a spin-ladder geometry.
The $a$-,$b$-, and $c$-axis is perpendicular to the ladder plane, along the leg, and along the rung direction, respectively
[see Fig.~1(a)]. Thus, ($bb$)[($cc$)] polarization denotes the leg (rung) polarization. Both (bb) and
($cc$) polarization are designated as intraladder polarization. (aa), (ab), and (ac) polarizations are collectively called
interladder polarizations.  The distinction between the intraladder and interladder polarization
allows separating intrinsic spin-ladder magnetic behavior from 3D effects.

\begin{figure}[htp]
\centering
\includegraphics[width=17cm]{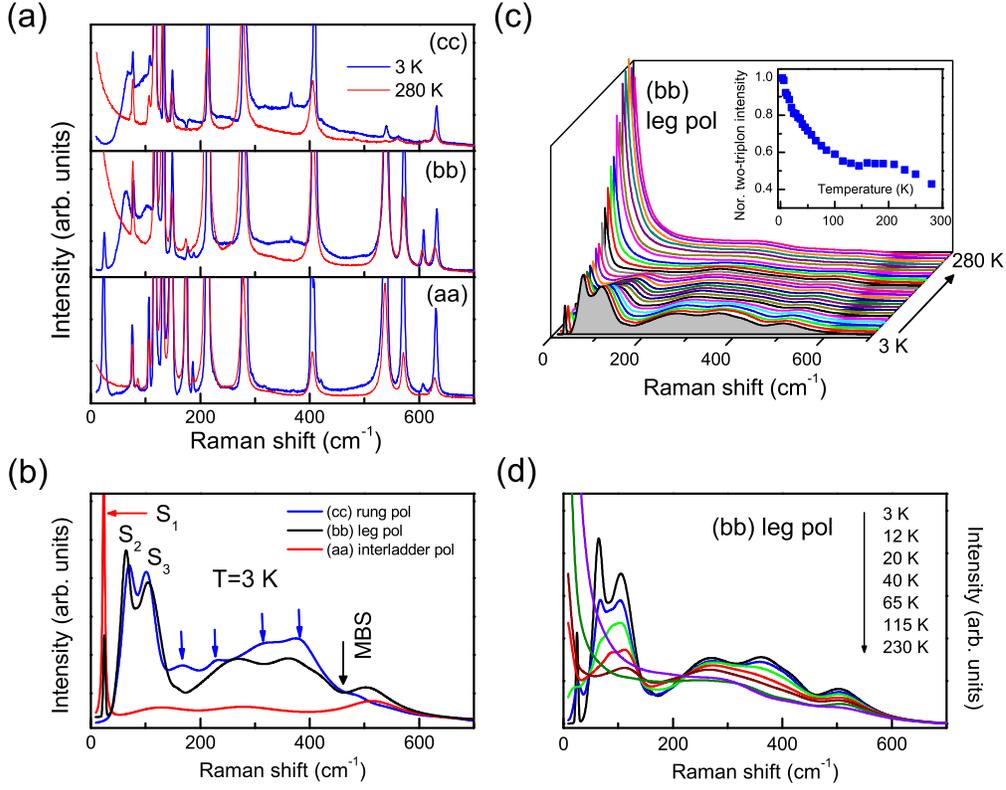}
\caption{(a) Comparison of Raman spectra
of BiCu$_2$PO$_6$ between 3~K and 280~K in ($cc$), ($bb$), and ($aa$) polarization. (b) Magnetic Raman scattering obtained in ($cc$), ($bb$), and ($aa$) polarization (after subtracting all phonon modes). The black arrow denotes a midband singularity (MBS) and S$_i$(i=1-3)
stands for singlet bound states. (c) Evolution of magnetic Raman spectra with increasing temperature in a 3D plot.
Inset: Temperature dependence of the two-triplon Raman scattering intensity. (d) Temperature dependence of the respective magnetic Raman spectra in a 2D plot.}
\label{FIG2}
\end{figure}

In Fig.~2(a) we observe a broad, structured  continuum at $T=3$~K in intraladder ($bb$) and ($cc$) polarizations. The continuum evolves to a quasi-elastic scattering at high temperatures. The temperature and polarization dependence of the continuum leads us to assign them to the above-discussed two-triplon excitations.

To allow a detailed study of the two-triplon energy, spectral weight, and shape, the phonons are subtracted. The polarization dependence of the resulting magnetic spectra at 3~K is presented in Fig.~2(b).
The broad magnetic continuum with a multiple-peak structure extends from 36~cm$^{-1}$ to 700~cm$^{-1}$.~\cite{KYC} For the leg ($bb$) and rung ($cc$) polarizations  the broad continuum and two resonance-like peaks at 64 (S$_2$) and 108~cm$^{-1}$ (S$_3$) (i.e. located above the two-particle threshold $2\Delta$) have complex lineshape with comparable intensity. In the interladder ($aa$) polarization the continuum is about four times weaker than in the intraladder polarization. The ratio of the integrated scattering intensity is $I^{(leg)}:I^{(rung)}:I^{(il)}=1:0.93:0.24$.
Since the magnetic Raman scattering intensity is given by $I \propto  |\langle i| S_i\cdot S_j |j\rangle|^2$, it scales in each polarization with the strength of exchange interactions. Thus, we can estimate the ratio of exchange constants using the relation $J^{(leg)}_{1}:J^{(rung)}_{4}:J^{(il)}_{3}=\emph{R}_{leg}\sqrt{I^{(leg)}}:\emph{R}_{rung}\sqrt{I^{(rung)}}:
\emph{R}_{il}\sqrt{I^{(il)}}$, where \emph{R} represents Raman scattering interactions.~\cite{Freitas}
Assuming that \emph{R} is proportional to the square of hopping matrix elements,~\cite{Tsirlin} we obtain $J^{(leg)}_{1}:J^{(rung)}_{4}:J^{(il)}_{3}=1:0.89:0.35$. This puts an upper limit to the strength of the interladder coupling due to the additional leg couplings, $J_2$ and $J'_2$, suggesting that the interladder coupling is substantial. Further, we note that the sharp peak at 24~cm$^{-1}$ (S$_1$) lying below the continuum is most pronounced in the interladder ($aa$) polarization. The three modes have ratios of $E^S_2/\Delta=1.01, 2.72, 4.59$. The polarization selection rule and energy scale indicate that the two S$_2$ and S$_3$ modes pertain to an intraladder physics while the S$_1$ mode is related to an interladder one.

Theoretical studies of two-leg ladders have shown that two-magnon bound states are always present in both the singlet and the triplet channel.~\cite{Uhrig,Damle,Sushkov,Kotov,Jurecka,Trebst,Zheng}
Based on the exact diagonalization of a realistic spin model for BiCu$_2$PO$_6$, Tsirlin {\it et al.}, also suggest the presence of a number of singlet bound states at the zone boundary.~\cite{Tsirlin}
We therefore assign the three modes at 24, 64, and 108 cm$^{-1}$ to anticipated singlet bound states. Below we will corroborate
this interpretation by analyzing the temperature and polarization dependence of them. Here we mention that
the resonance-like behavior of the bound modes points to their singularity-like nature and
is attributed to a strong attractive interaction of neighboring triplets due to in-leg frustration.

We recall that the singlet bound state has been observed in the two-leg compound Ca$_{14-x}$La$_x$Cu$_{24}$O$_{41}$ with $J_\perp/J_\|\approx 1$ via phonon-assisted two-magnon absorption.~\cite{Windt}
The two-magnon singlet bound state has a two-peak feature assigned to van-Hove singularities in the density of states of the singlet dispersion at the Brillouin zone boundary and at $k\approx \pi/2$.
The spectral weight and shape of the two-magnon-plus-phonon absorption spectrum bears a close resemblance to our two-triplon Raman spectra [compare Fig.~3 of Ref.~\cite{Windt}
and Fig.~2(a)]. In spite of the apparent similarity, however, the number and details of the singlet bound states differ, pointing to a significance of frustration and enlarged exchange degrees of freedom in BiCu$_2$PO$_6$. In contrast to the optical conductivity, Raman scattering measurements probe magnetic excitations without involving a phonon. Thus, the three peaks (S$_1$, S$_2$, and S$_3$) should be interpreted as separate bound states.

We now turn to the two-triplon continuum at higher energies. The Raman response of the two-leg ladder was calculated by using the continuous unitary transformation method.~\cite{Schmidt,Knetter,Schmidt05}
For a finite rung coupling, the main feature is the existence of a midband singularity(MBS), which  splits the continuum into two bands. This is attributed to a combined effect of a van-Hove singularity in the triplet-dispersion near $k=0$ and the orthogonal relation between excited triplets induced by triplet-triplet interactions. This spin ladder feature is not visible in Ca$_{14-x}$La$_x$Cu$_{24}$O$_{41}$ due to sizable four-spin cyclic exchange interactions.~\cite{Sugai,Gozar}
On the contrary, the studied compound exhibits a MBS at 455~cm$^{-1}$, which separates the secondary band at about 510~cm$^{-1}$ from the main band centered around 300~cm$^{-1}$ [see the black arrow in Fig.~2(b)]. This confirms that the magnetic excitations {\it at high energies} are largely described by isolated ladder-like physics. We remark that the simple spin ladder model also reproduces the magnetic behavior upon chemical substitution.~\cite{Bobroff,Casola} However, a closer look unveils a fine structure with multiple maxima at 254, 366, and 418~cm$^{-1}$ in the leg ($bb$) polarization and 110, 229, 310, 377, and 439~cm$^{-1}$ in the rung ($cc$) polarization [see the arrows in Fig.~2(b)]. The strong polarization dependence of the spectral shape and the presence of a number of peaks within the continuum are beyond the spin ladder physics. Kohama {\it et al.} reached essentially the same conclusion based on the field-induced anisotropic, complex phase diagram.~\cite{Kohama}  We ascribe
the fine features of the continuum to high-energy bound modes and/or van-Hove singularities related
to the frustrated spin chain.  Here we summarize the multi-facetted magnetic behavior of BiCu$_2$PO$_6$ according to an energy hierarchy.  At high energies the magnetic behavior approaches that of an isolated spin ladder as demonstrated by the MBS. At intermediate energies the intraladder bound modes show up as frustrated 1D effects while at low energies 3D interladder interactions lead to the interladder bound mode.

\begin{figure}[htp]
\centering
\includegraphics[width=11cm]{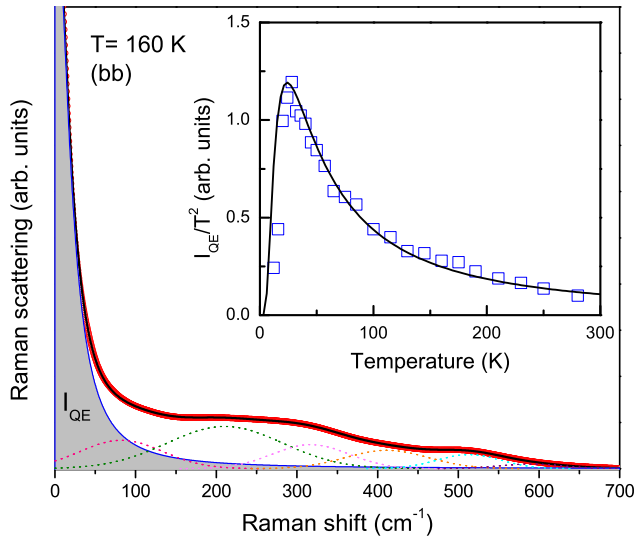}
\caption{Representative magnetic continuum at 160~K and in ($bb$) polarization, which is fitted to a sum of one Lorentzian
and six Gaussian profiles. The gray Lorentzian spectrum denotes the quasielastic (QE) scattering response.
Inset: The intensity of the quasielastic scattering is normalized by $T^2$ to be compared with the magnetic
specific heat ($C_m$). The solid line indicates the calculated zero-field specific heat.}
\label{FIG3}
\end{figure}

In Figs.~2(c) and (d) the detailed temperature dependence of the magnetic continuum in the leg polarization is plotted. With increasing temperature the low-energy singlet bound modes dissolve into a continuum of excitations and then into quasi-elastic scattering upon further heating. This is easily understood as a thermal destruction of bound states.
We note that essentially the same temperature dependence of the singlet bound states is observed in the frustrated, two-dimensional Shastry-Sutherland system SrCu$_2$(BO$_3$)$_2$.~\cite{Lemmens00} In this system the multiple bound states are due to the localization of triplet excitations on a strongly frustrated lattice.

With increasing temperature the two-triplon continuum is systematically suppressed while the spectral weight shifts toward lower energies. The integrated intensity of the magnetic continuum, corrected by the Bose-Einstein factor, is shown as a function of temperature in the inset of Fig.~2(c). Upon cooling, the intensity increases monotonically and shows no saturation towards lower temperatures. This is due to a thermal population of triplets as found in other spin-gapped systems.~\cite{Choi05} Despite the apparent resemblance, there is some discrepancy. In case of the spin gap systems, the magnetic continuum is gradually suppressed without appreciable changes in its spectral shape. Therefore, the softening continuum observed in BiCu$_2$PO$_6$ might be due to the damping of high-energy bound modes which lie within the background continuum and appear as multiple maxima.

We will now turn to the low-energy quasielastic (QE) scattering, which provides information about the critical spin dynamics. In low-dimensional antiferromagnets, the QE scattering is due to fluctuations of the magnetic energy density;~\cite{Choi11}
\begin{equation}
I_{QE}(\omega) \propto \int^{\infty}_{-\infty} e^{-i\omega t}dt \langle E(k,t)E^*(k,0)\rangle,
\end{equation}
where $E(k,t)$ is the Fourier transform of $E(r)=-\langle \sum_{i>j}J_{ij}S_i\cdot S_j\delta(r-r_i)\rangle$.
Applying the fluctuation-dissipation theorem in the hydrodynamic and high-temperature limit, Eq.~(1) is simplified to a Lorentzian
profile,
\begin{equation}
I_{QE}(\omega) \propto C_m T^2 \frac{Dk^2}{\omega^2+(Dk^2)^2},
\end{equation}
where $C_m$ is the magnetic specific heat and $D$ is the thermal diffusion constant.
Eq.~(2) allows deriving $C_m$  by exploiting the relation $C_m(T)\propto I_{QE}(T)/T^2$.
For a quantitative analysis we fit the magnetic scattering to a sum of Eq.~(2) (solid line) and Gaussian profiles (dashed lines). The representative fit is displayed in Fig.~3. The gray area, described by the Lorenztian profile,
corresponds to $I_{QE}$. In the inset of Fig.~3 we show the temperature dependence of the derived $C_m(T)$. It shows a Schottky-type maximum at about 25~K, which is characteristic for a spin-gap system. The whole temperature dependence is fitted to an expression of $C_{m}\propto T^{-3/2}\exp(-\Delta/T)$ to extract a spin gap.~\cite{Hagiwara}
The fitting result is depicted as the solid line with the the spin gap of $\Delta=36$~K. The obtained value is very close to the spin gap determined by thermodynamic measurements.

\begin{figure}[htp]
\centering
\includegraphics[width=13cm]{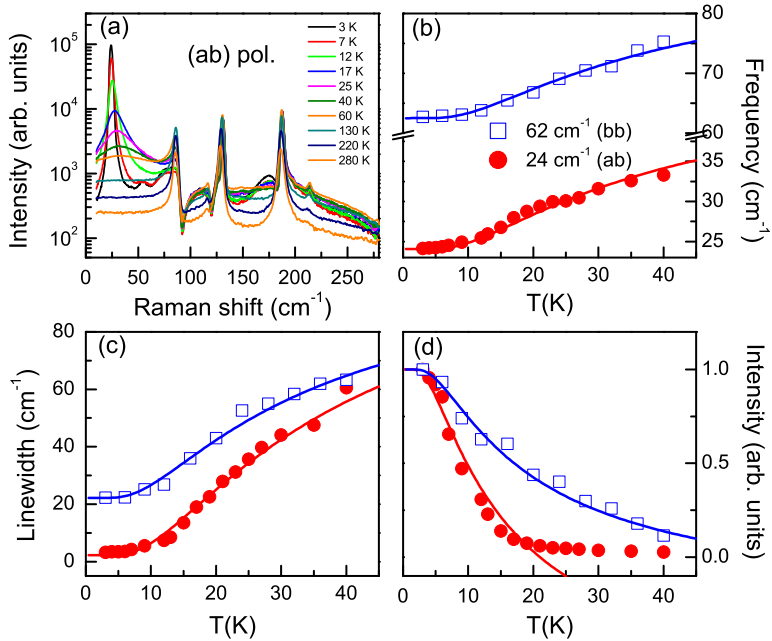}
\caption{(a) Temperature dependence of Raman spectra in interladder ($ab$) polarization on a logarithmic intensity scale. Temperature dependence of (b) frequency, (c) linewidth, and (d) normalized intensity of the singlet bound state at 24~cm$^{-1}$ in interladder ($ab$) polarization (full circles) and at 62~cm$^{-1}$ in leg ($bb$) polarization (open squares). The solid lines are fits to the data, see the text for details.}
\label{FIG4}
\end{figure}

Figure 4(a) shows the temperature dependence of Raman spectra in interladder ($ab$) polarization on a logarithmic intensity scale. We identify a resonance-like singlet bound mode at $E^S_2=24~\mbox{cm}^{-1}$. The singlet bonding energy is estimated to $\epsilon^S=2\Delta-E^S_2=23~\mbox{cm}^{-1}$. The lowest singlet bound mode shows up in all polarizations except the leg ($bb$) polarization. The scattering intensity is most pronounced in ($ab$) and ($aa$) polarizations, that is, in interladder scattering geometry. Since in the exchange light scattering mechanism the Raman operator is identical to the spin Hamiltonian, the intensity of the bound states should be observed with polarizations parallel to the dominant exchange path, i.e., along the leg and the rung direction. Therefore, the polarization selection rule indicates a substantial interladder coupling and the off-diagonal scattering matrix element, indicative of strong anisotropic exchange interactions. Here we note that the higher-energy singlet modes, S$_2$ and S$_3$ are observed only in intraladder polarizations, confirming that they are part of an isolated spin ladder. The intraladder singlet modes are interpreted as a softening singlet mode, which results from the transition from a dimerized to a spin ladder phase.~\cite{Vekua}
The melting of the dimer crystal corroborates that BiCu$_2$PO$_6$ has a RVB ground state while its spin dynamics is governed by singlet fluctuations. Therefore, BiCu$_2$PO$_6$ lies in the vicinity to a dimer phase. This is further supported by the numerical calculations that show a crossover from a spin chain to a spin ladder~\cite{Tsirlin} and provides a convincing explanation for the field-induced anisotropic commensurate-incommensurate transition.~\cite{Kohama}
In addition, the interladder bound mode can be taken as an indication to the proximity to a 3D long-range ordered state
from the RVB one. Future pressure work is required to test this possibility.

The temperature dependence of the frequency, linewidth, and normalized intensity of the interladder and intraladder bound modes is compared in Figs.~4(b), (c), and (d). With increasing temperature both bound modes undergo a large hardening by 10-12~cm$^{-1}$ toward higher frequencies and broaden enormously. At the same time, their scattering intensity decreases rapidly. Since the singlet bound states are susceptible to scatterings on thermally excited triplons, the reduction of the bound energy, the damping, and the decrease of the scattering intensity with temperature are related to the density of triplet excitations. The parameters of the bound modes are fitted to a single function, $1-A \exp(-\Delta_S/k_BT)$ where $A$ is a constant. The frequency shift and the damping are reasonably well described by the above equation with $\Delta_S= 30-35$~K. However, sizable deviation with a much smaller $\Delta_S=11-14$~K is seen in the temperature dependence of the intensity. This might be due to a broad triplon band and/or proximity to a quantum critical point.

In conclusion, we have presented a comprehensive Raman scattering study of the magnetic excitation spectrum in BiCu$_2$PO$_6$, a frustrated spin-ladder compound with interladder couplings. We observe the presence of both interladder and intraladder singlet bound modes at low energies as well as of a broad continuum of two-triplet excitations. The intraladder singlet bound states are a consequence of the melting of a dimer crystal into a RVB state  and thus prove a Z$_2$ transition between two different singlet states. The interladder singlet bound state indicates an instability toward another ordered phase.
BiCu$_2$PO$_6$ is a model system, which hosts two different kinds of quantum phase transitions in a single system.

We thank W. Brenig for insightful discussions.
KYC acknowledges financial support from the Alexander-von-Humboldt Foundation and Korea NRF Grant (No. 2010-0011325). FCC acknowledges
support from NSC-Taiwan under project number NSC 100-2119-M-002-021.
GJS acknowledges support from NSC-Taiwan under project number NSC
100-2112-M-002-001-MY3.

%\bibliography{achemso}

\begin{thebibliography}{99}
\bibitem{KYC}
In magnetic Raman scattering, the onset energy corresponds to twice the spin gap, yielding $\Delta_R=18$~cm$^{-1}(\simeq 26$~K). The value is rather small compared to the spin gap of $\Delta=32$~K determined by high-field magnetization measurements. The difference is due to the vicinity of singlet bound states to the spin gap.
\end{thebibliography}

\end{document}